\documentclass[twocolumn,aps,pra,superscriptaddress,10pt]{revtex4-2}
\usepackage{graphicx}
\usepackage{amsmath}
\usepackage{amssymb}
\usepackage{braket}
\usepackage{bm}
\usepackage{dcolumn}
\usepackage{subfigure}
\usepackage{graphicx}
\usepackage[colorlinks=true,linkcolor=blue,citecolor=blue,urlcolor=blue]{hyperref}
\usepackage{soul}
\usepackage{changes}
\usepackage{amsthm,amsmath,amssymb}

\usepackage{mathrsfs}
\usepackage{multirow}

\usepackage[percent]{overpic}

\makeatother

\begin{document}

\title{Manipulation of Spin-Exchange Interaction Between Ultracold Alkaline-Earth Atoms via Shaking Lattice}

\author{Chen Xu}

\affiliation{School of Physics, Renmin University of China, Beijing, 100872,
China}

\author{Peng Zhang}
\email{pengzhang@ruc.edu.cn}

\affiliation{School of Physics, Renmin University of China, Beijing, 100872,
China}

\affiliation{Key Laboratory of Quantum State Construction and Manipulation (Ministry of Education), Renmin University of China, Beijing, 100872, China}

\date{today}

\begin{abstract}
We propose a method to modulate the spin-exchange interaction between ultracold alkaline-earth atoms in distinct electronic orbital states via periodic modulation of the optical lattice depth. Unlike previous techniques relying on confinement-induced resonances, our approach {\it does not require} the interatomic $s$-wave scattering length to be comparable to the characteristic length of the confinement potential. Consequently, it is broadly applicable to various alkaline-earth species, particularly those lacking very-large scattering lengths, {\it e.g.}, $^{171}$Yb. Our method is helpful for the quantum simulation of many-body phenomena associated with local magnetic impurities, such as the Kondo effect.

%
%

\end{abstract}
\maketitle

\section{Introduction}

\begin{figure}[t]
\includegraphics[width=0.7\linewidth]{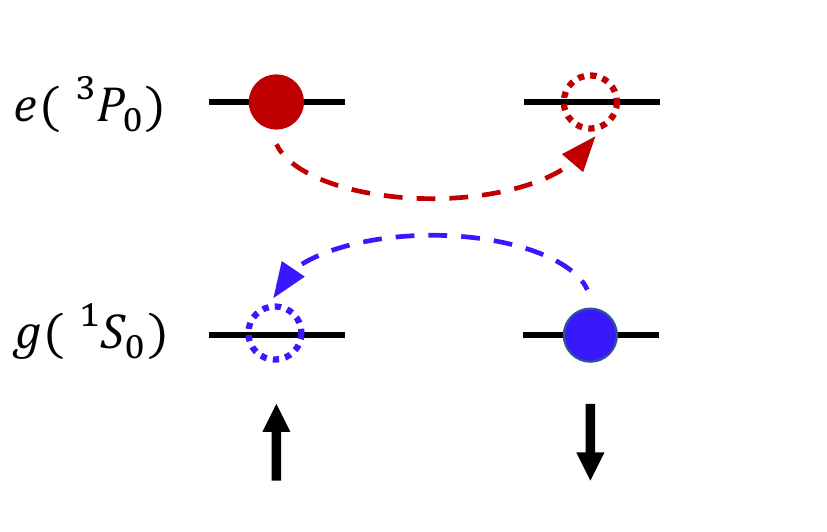}
	\caption{Schematic illustration of the spin-exchange interaction (SEI)
		between alkaline-earth-metal atoms in the $^{1}$S$_{0}$ ($g$) 
		and $^{3}$P$_{0}$ ($e$) states. The nuclear-spin states are denoted 
		as $\uparrow$ and $\downarrow$.
		}
	\label{scheme}
\end{figure}

In recent years, ultracold alkaline-earth(-like) atoms have attracted considerable attention as a promising platform for simulating phenomena induced by magnetic impurities, such as the Kondo effect. Representative atomic species include
 $^{171}$Yb, $^{173}$Yb, and $^{87}$Sr
\cite{PengZhang2016PRA, PengZhang2017PRA, PengZhang2018PRA, PengZhang2020PRA, PengZhang2020Review, PengZhang2015PRL, Fallani2015PRL, Folling2015PRL, Folling2019PRL, Riegger2018PRL, Takahashi2019PRA, Takahashi2021PRA, YeJun2014Science}. These atoms can be prepared in either the $^{1}$S$_{0}$ ground electronic state ($g$) or the long-lived $^{3}$P$_{0}$ metastable state ($e$). Additionally, a nuclear spin-exchange interaction (SEI) naturally arises between atoms in the $g$- and $e$-states (Fig.~\ref{scheme}), providing a direct analog to the spin-exchange coupling in solid-state materials. Consequently, when the $e$-state atoms are tightly confined in local trapping potentials while the $g$-state atoms remain itinerant, they can effectively emulate localized magnetic impurities and conduction electrons, respectively.

To realize the aforementioned quantum simulation, it is necessary to develop techniques to efficiently manipulate the SEI between the $g$- and $e$-atoms. Unfortunately, one cannot tune the SEI via magnetic-field induced Feshbach resonances \cite{PengZhang2015PRL, Fallani2015PRL, Folling2015PRL, Folling2019PRL}, as is commonly done for ultracold alkali atoms. That is because such an approach would lead to serious heating effects, since the nuclear-spin states before and after the SEI process would possess different Zeeman energies under an applied magnetic field. Therefore, alternative methods for manipulating the SEI must be developed.

One such approach uses  confinement-induced resonances (CIRs) to manipulate the SEI \cite{PengZhang2016PRA, PengZhang2017PRA, PengZhang2018PRA, PengZhang2020PRA, PengZhang2020Review, Riegger2018PRL}. This approach enables resonant tuning of the SEI by varying the confinement strength or the optical-lattice depth. However, to date, this manipulation technique has only been experimentally realized for $^{173}$Yb atoms, and it is difficult to implement for other types of fermionic alkaline-earth-like atoms, such as $^{171}$Yb or $^{87}$Sr. This difficulty is due to the following fact. The occurrence of a CIR requires the interatomic scattering length between the $g$- and $e$-atoms to be comparable to or larger than the characteristic length of the optical confinement potential, which is on the order of $1000a_0$ in typical experiments, where $a_0$ is the Bohr radius. While this condition is readily satisfied by $^{173}$Yb atoms, the relevant scattering lengths for $^{171}$Yb and $^{87}$Sr atoms are significantly less than $1000a_0$ \cite{PengZhang2022PRR,YeJun2014Science}, making CIRs highly impractical to realize in these systems.

In this work, we propose another approach to manipulate the SEI between a mobile $g$-atom in a quasi one-dimensional (quasi 1D) optical lattice and an $e$-atom localized at a particular site. This method does not impose stringent requirements on the inter-atomic scattering length, making it broadly applicable to various atomic species. Specifically, in our approach the manipulation of the SEI is achieved via a three-color periodic modulation (shaking) of the optical lattice strength. Our method is a generalization of the approach for manipulating interactions between ultracold atoms in a 1D lattice, which were proposed by L. Cardarelli {\it et al.} \cite{Santos2016PRA}, and was recently applied in the experimental realization of 1D anyons using ultracold $^{87}\mathrm{Rb}$ atoms \cite{Greiner2024Science}. Taking $^{171}$Yb atoms as an example, we  estimate the experimental parameters required to implement our approach. The physical effects of this SEI manipulation are further illustrated by calculating the spin-exchange scattering probability between the $g$- and $e$-atoms.

The remainder of this paper is organized as follows.
 In Sec.~\ref{3cm} we show the principle of our approach for the manipulation of SEI via shaking lattice. The experimental conditions for realizing our scheme are estimated in Sec.~\ref{exp} for $^{171}$Yb atoms. In Sec.~\ref{scattering}, we calculate the interatomic spin-exchange probability for our system. Finally, Sec.~\ref{summary} provides a summary. Some details of our calculations are given in the appendixes. 



\section{Manipulation of SEI via Shaking Lattice}
\label{3cm}

\subsection{Setup}
\label{setup}

Let us consider the quasi-1D system shown in Fig.~\ref{scheme2}(a).
Specifically,  all $g$- and $e$-atoms, each has nuclear spin states $\uparrow$ or $\downarrow$,
 are confined in the same quasi-1D trapping potential (2D harmonic potential in the $x$-$y$ plane). 
This 
quasi-1D confinement
 is generated by laser beams with a magic frequency, for which the $g$- and $e$-atoms have the same AC polarizability. 
Note that the axial direction ($z$-direction) of this quasi-1D confinement is tilted with respect to the horizontal plane.
Moreover, there is a state-dependent 1D optical lattice along the $z$-direction, which is created by  laser beams
with another frequency.
This 1D optical lattice potential is shallow and deep for the $g$- and $e$-atoms, respectively. As a result, 
each $e$-atom is localized at an individual site, and all the  $g$-atoms can hop between nearest-neighbor sites.

Moreover, the  motion of the atoms in the $x$-$y$ plane are frozen in the ground state of the 
two-dimensional harmonic confinement.
As a result, only the motion of the $g$-atom in the 1D lattice along the $z$-direction, and the evolution of the 
spin states of the $g$- and $e$-atoms, are required to be studied. 
Furthermore, the strength of this lattice is shaking, so that the hopping rate of the $g$-atom between nearest neighbor sites is modulated. As shown below, the effective SEI is manipulated
by the amplitude and frequencies of this modulation.

 \subsection{Two-Body Effective Hamiltonian}
\label{tb}


\begin{figure}[t]
	\centering
	
	\begin{overpic}[width=0.7\linewidth]{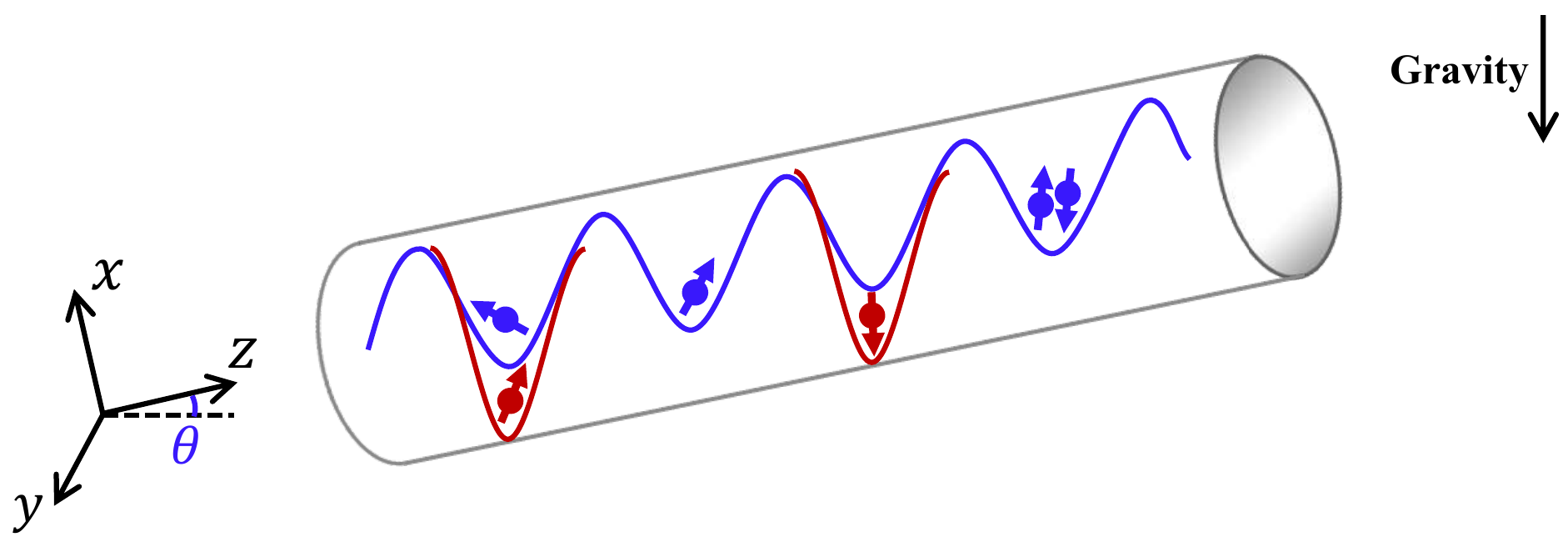}
		\put(10,40){\bfseries (a)}
	\end{overpic}
	
	\vspace{1.9em}
	
	\begin{overpic}[width=0.6\linewidth]{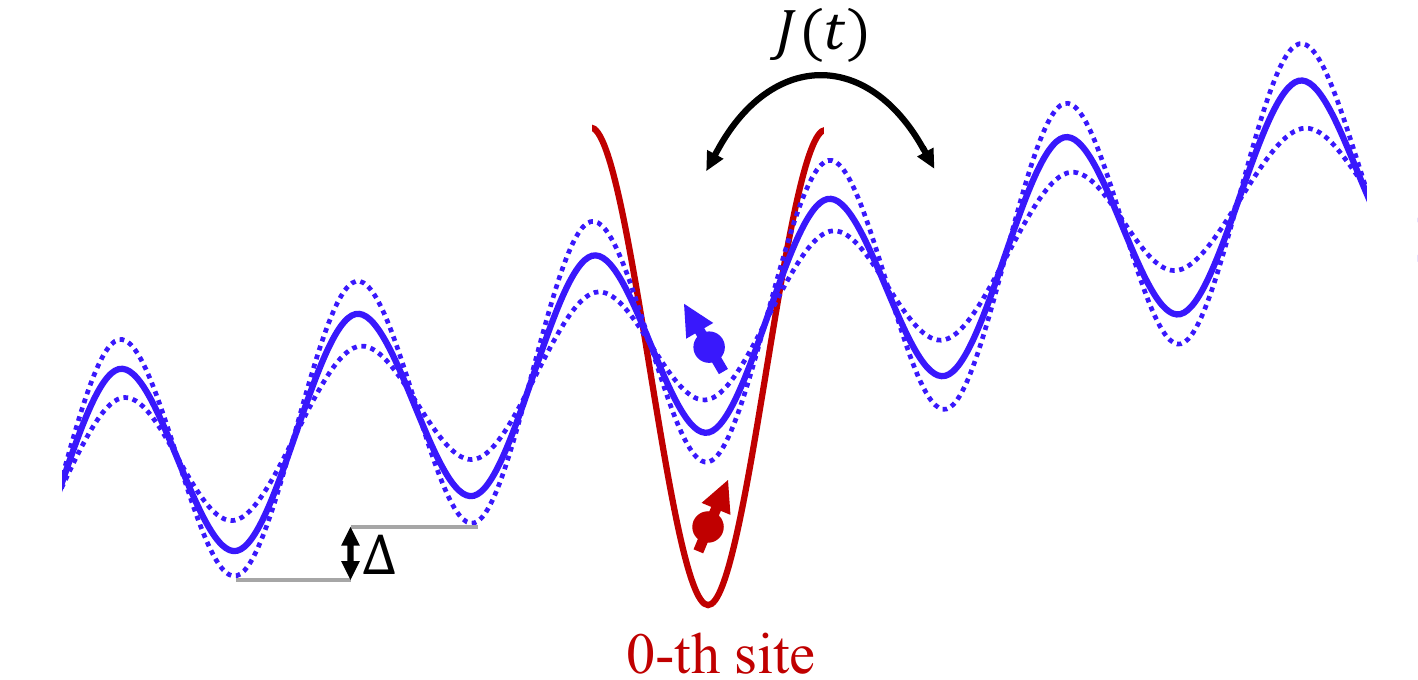}
		\put(2,40){\bfseries (b)}
	\end{overpic}
	
	\vspace{2.5em}
	
	\begin{overpic}[width=0.6\linewidth]{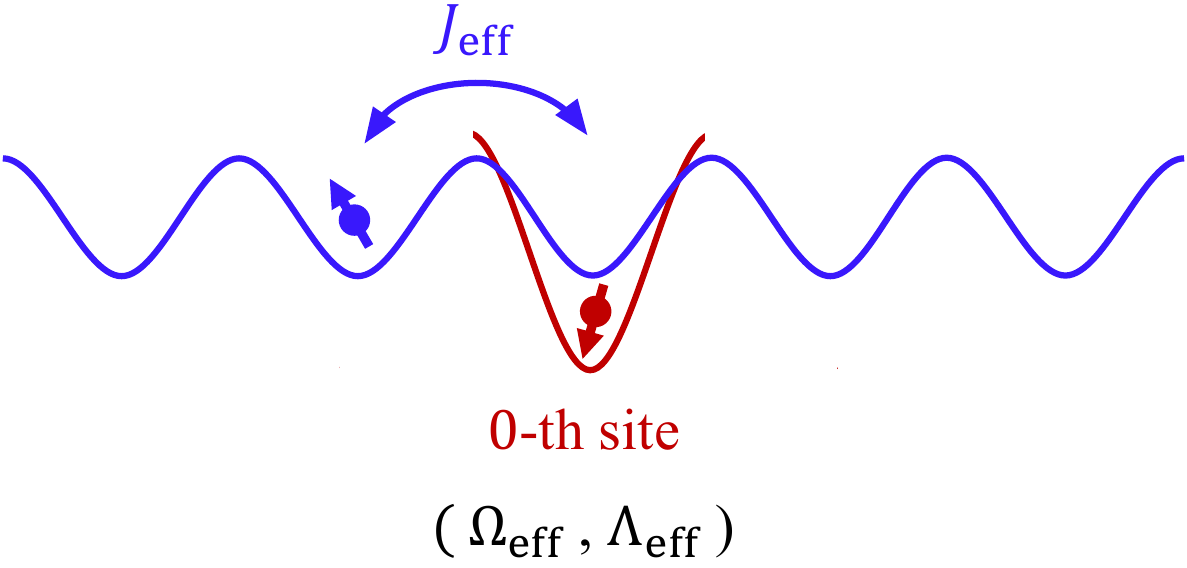}
		\put(2,40){\bfseries (c)}
	\end{overpic}
	
	\vspace{2.5em}
	
	\begin{overpic}[width=0.6\linewidth]{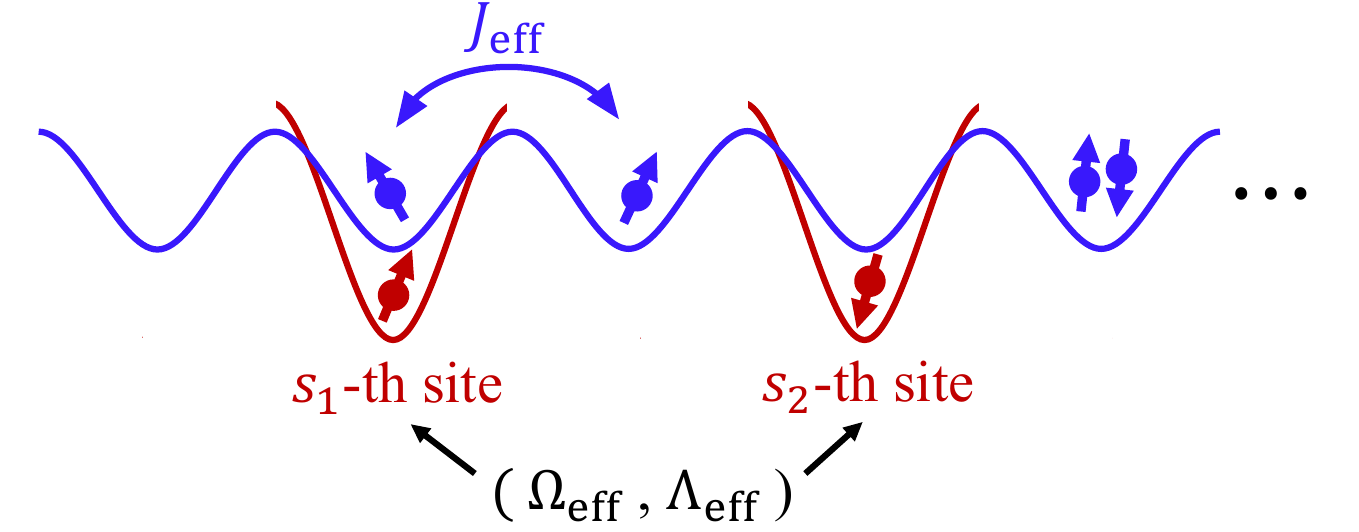}
		\put(2,40){\bfseries (d)}
	\end{overpic}
	
	\caption{{\bf (a):} The quasi-1D system of our approach.
		The tilted quasi-1D confinement is created by laser beams with magic frequency, and the state-dependent 1D lattice along the $z$-direction is created by beams with other frequency. 
		{\bf (b):} The tilted 1D shaking lattice model for one $g$-atom and one $e$-atom. 
		{\bf (c):} The effective 1D lattice model realized by our approach, for one $g$-atom and one $e$-atom.
		{\bf (d):} The effective 1D lattice model realized by our approach, for many-body cases.}
	\label{scheme2}
\end{figure}

To illustrate our scheme in detail, here we start from a simple case with one $g$-atom and one $e$-atom. As shown above, the $e$-atom is located at a particular site ($0$-th site) of the 1D tilted lattice along the $z$-direction (Fig.~\ref{scheme2}(b)). Taking into account the shaking of the lattice and the SEI between these two atoms, we can express the Hamiltonian of the spatial motion of the $g$-atom and the spin of all the two atoms as
\begin{eqnarray}
	&&\hat H(t)=\nonumber\\
	&&\bigg[-J(t)\sum_{j=-\infty}^{+\infty}\big(|j+1\rangle_g \langle j|+h.c.\big)+\Delta\sum_j j |j\rangle_g \langle j|\bigg]\otimes \hat {\cal I}_{\rm spin}\nonumber\\
	&&+|j=0\rangle_g \langle j=0|\otimes\left(U_{+}\hat{\cal P}_++U_-\hat{\cal P}_-\right).\label{Ht}
\end{eqnarray}
Here
 $|j\rangle_g$ ($j=0,\pm 1, \pm 2,...$) is the  state of the $g$-atom, which means that
 this atom is located at the $j$-th site of the lattice, and $\hat {\cal I}_{\rm spin}$ is the identity operator of the 
 Hilbert space for the spin of both two atoms. Additionally, in Eq.~(\ref{Ht}), the positive parameter
  $\Delta$ is the slope of the tilted lattice, and $J(t)$ is the hopping rate of the $g$-atom, which  is time-dependent due to the shaking of the lattice potential.
 Furthermore, the inter-atomic SEI  is described by the term $|j=0\rangle_g \langle j=0|\otimes(U_{+}\hat{\cal P}_++U_-\hat{\cal P}_-)$ of 
 Eq.~(\ref{Ht}), where 
 $\hat{\cal P}_{+(-)}$ are the projection operators onto the spin singlet (triplet) states. Specifically, we have 
\begin{eqnarray}
\hat{\cal P}_{+}&=&|{\rm S}\rangle\langle{\rm S}|;
\ \ \ \ 
\hat{\cal P}_{-}=
\sum_{m=0,\pm 1}|{\rm T},m\rangle\langle{\rm T},m|,\label{proj}
\end{eqnarray}
with
\begin{eqnarray}
|{\rm S}\rangle&=&\frac{1}{\sqrt{2}}\Big(|\uparrow\rangle_g|\downarrow\rangle_e-|\downarrow\rangle_g|\uparrow\rangle_e\Big)\label{singlet}
\end{eqnarray}
being the singlet state, and 
\begin{eqnarray}
|{\rm T},-1\rangle&=&|\downarrow\rangle_g|\downarrow\rangle_e;\label{t1}\\[8pt]
|{\rm T},0\rangle&=&\frac{1}{\sqrt{2}}\Big(|\uparrow\rangle_g|\downarrow\rangle_e+|\downarrow\rangle_g|\uparrow\rangle_e\Big);\label{triplet0}\\[8pt]
|{\rm T},1\rangle&=&|\uparrow\rangle_g|\uparrow\rangle_e\label{t3}
\end{eqnarray}
being the triplet states.
Moreover, $U_{+(-)}$  is the interaction strength between the $g$ and $e$ atoms, which are both in the $0$-th site and in the nuclear spin singlet (triplet) states.

 Clearly, the values of $J(t)$, $\Delta$, and $U_{\pm}$ in our model of Eq.~(\ref{Ht}) are determined by original parameters of the realistic experiments, such as the strength of the  laser used to create the lattices,  and the inter-atomic scattering lengths  in the 3D free space. In the next section we will discuss the required values of these experimental parameters in detail.  
 Here we illustrate our approach for the manipulation of SEI via discussions based on the Hamiltonian $\hat H(t)$ of Eq.~(\ref{Ht}).

 In our approach, a three-color periodic modulation is performed to the hopping rate $J$. Under this modulation, we have
\begin{eqnarray}
	J(t) = J_{0} +\delta J\sum_{l=1}^3 \bigg(e^{i\omega_lt}+ e^{-i\omega_lt}\bigg),\label{Jt}
\end{eqnarray}
where 
\begin{eqnarray}
\omega_1=\Delta;\ \ 
\omega_2=\Delta+\nu;\ \ \omega_3=\Delta-\nu, \label{fre}
\end{eqnarray}
with the positive parameter $\Delta$ being just the 
aforementioned slope of the tilted lattice, and $\nu>0$. In our scheme
$\{J_0, \delta J,\nu\}$ are control parameters.

Now we derive the effective Hamiltonian for our system by generalizing the derivation of Ref.~\cite{Santos2016PRA} to the current case. To this end, 
we introduce a rotating frame $R$. Specifically, the quantum state $|\psi(t)\rangle_R$ in this frame is related to the state  $|\psi(t)\rangle_S$ in the Schr\"odinger picture via
\begin{eqnarray}
|\psi(t)\rangle_R=e^{i\hat H_0t}|\psi(t)\rangle_S,\label{rfr}
\end{eqnarray}
where 
\begin{eqnarray}
	\hat H_0&=&\bigg[\Delta\sum_j j |j\rangle_g \langle j|+\nu |j=0\rangle_g \langle j=0|\bigg]\otimes {\hat {\cal I}_{\rm spin}}.\label{H0}
\end{eqnarray}
In this rotated frame, the Hamiltonian of our system is 
\begin{eqnarray}
\hat H_R(t)=e^{i\hat H_0t}\big[\hat H(t)-\hat H_0\big]e^{-i\hat H_0t}.\label{hrt}
\end{eqnarray}
Substituting Eqs.~(\ref{Ht}, \ref{Jt}, \ref{H0}) into Eq.~(\ref{hrt}), we can directly obtain the expression of $\hat H_R(t)$, which includes both time-independent terms and oscillating terms. Furthermore, under the conditions
\begin{eqnarray}
J_0, \ \delta J&\ll&
\Delta,\ \nu, \ U_\pm, \ |2\Delta-\nu|,\ |\Delta-U_\pm|, \ |2\Delta-U_\pm|,\nonumber\\
&&|2\Delta-\nu+U_\pm|,\ |2\Delta-\nu-U_\pm|,\ |2\Delta+\nu-U_\pm|,\nonumber\\
\label{c3}
\end{eqnarray}
all the oscillating terms of $\hat H_R(t)$ can be ignored (Appendix~\ref{condition}). Thus,
the effective Hamiltonian in the rotating frame $R$ is just
\begin{eqnarray}
\hat H_{\rm eff}&=&\bigg[-J_{\rm eff}\sum_{j=-\infty}^{+\infty}\big(|j+1\rangle_g \langle j|+h.c.\big)\bigg]\otimes {\hat {\cal I}_{\rm spin}}\nonumber\\
[3pt]
	&&+|j=0\rangle_g \langle j=0|\otimes\left(U^{\rm (eff)}_{+}\hat{\cal P}_++U^{\rm (eff)}_-\hat{\cal P}_-\right),\nonumber\\
\label{heff1}
\end{eqnarray}
where
\begin{eqnarray}
J_{\rm eff}=\delta J,\hspace{0.6cm}
U^{\rm (eff)}_{\pm}=U_{\pm}-\nu.\label{jueff}
\end{eqnarray}
Note that the effective Hamiltonian (\ref{heff1}) can be directly re-written as 
\begin{eqnarray}
\hat H_{\rm eff}&=&\bigg[-J_{\rm eff}\sum_{j=-\infty}^{+\infty}\big(|j+1\rangle_g \langle j|+h.c.\big)\bigg]\otimes {\hat {\cal I}_{\rm spin}}\nonumber\\
[3pt]
&&+|j=0\rangle_g \langle j=0|\otimes\bigg[\Lambda_{\rm eff}
+\Omega_{\rm eff}\hat{\bm \sigma}^{(g)} \cdot \hat{\bm \sigma}^{(e)}\bigg],\nonumber\\
\label{heff2}
\end{eqnarray}
where $\hat{\bm \sigma}^{(g/e)}=(\hat\sigma_x^{(g/e)},\hat\sigma_y^{(g/e)},\hat\sigma_z^{(g/e)})$ are the Pauli-operator vectors for the nuclear spins of the $g/e$-atom, and the parameters $\Omega_{\rm eff}$ and $\Lambda_{\rm eff}$ are defined as
\begin{eqnarray}
\Omega_{\rm eff}=\frac{U_{-}^{(\rm eff)}-U_{+}^{(\rm eff)}}{4}\label{Omega};\hspace{0.5cm}
\Lambda_{\rm eff}=\frac{3U_{-}^{(\rm eff)}+U_{+}^{(\rm eff)}}{4},\label{Lambda}
\end{eqnarray}
with $J_{\rm eff}$ and  $U_{\pm}^{(\rm eff)}$ being given by Eq.~(\ref{jueff}).

In summary, as illustrated in Fig.~\ref{scheme2}(c), the itinerant $g$-atom effectively moves in an untilted lattice with hopping rate $J_{\rm eff}$. Its interaction with the localized $e$-atom is described by the effective parameters $U_{\pm}^{(\rm eff)}$, or equivalently, $(\Omega_{\rm eff},\Lambda_{\rm eff})$.

\subsection{Many-Body Effective Hamiltonian}
\label{mb}

Now let us consider 
the general case, with many $g$- and $e$-atoms being in our tilted quasi-1D lattice. As above, we assume that the $e$-atom  $1,2,...,N_e$ are confined in 
$s_1,s_2,...,s_{N_e}$ sites, respectively, and the $g$-atoms can hop between nearest neighbor sites. Our above two-body results can be straightforwardly generalized to this case. Specifically,
when the hopping rate $J$ is modulated as in Eq.~(\ref{Jt}), and  the condition (\ref{c3}) is satisfied, the effective Hamiltonian in the rotating frame can be expressed in the second-quantization  form as
\begin{eqnarray}
	&&\hat H_{\rm  MB}^{\rm  (eff)}=\nonumber\\
	&&-J_{\rm eff}\sum^{\infty}_{j=-\infty}\sum_{\sigma=\uparrow,\downarrow}\left(\hat c^{\dagger}_{j+1,\sigma}\hat c_{j,\sigma}+{\rm  h.c.}\right)\nonumber\\
	&&+\Omega_{\rm eff}\sum_{\alpha=1}^{N_e}\left(
	\hat{\bm \sigma}^{(g,s_\alpha)} \cdot \hat{\bm \sigma}^{(e,\alpha)}\right)
	+\Lambda_{\rm eff} \sum_{\alpha=1}^{N_e} \hat  n^{(g,s_\alpha)},\label{hmb1}
\end{eqnarray}
where 
$\Omega_{\rm eff}$  and  $\Lambda_{\rm eff}$ are given by Eq.~(\ref{Omega}),
$\hat c_{j,\sigma}$  and  $\hat c^\dagger_{j,\sigma}$  ($\sigma=\uparrow,\downarrow$)  are  the  annihilation  and  creation  operators  of  $g$-atom  at  the  $j$-th  site  of  the  axial  lattice  with  nuclear  spin  $\sigma$,  respectively,  and  the Pauli-operator vectors  $\hat{\bm \sigma}^{(g,j)}=\left(\hat\sigma_x^{(g,j)},\hat\sigma_y^{(g,j)},\hat\sigma_z^{(g,j)}\right)$  and  the density operators $\hat n^{(g,j)}$ ($j=0,\pm 1,\pm 2,...$) are  defined  as
\begin{eqnarray}
	\hat \sigma_{x}^{(g,j)}&=&\hat c^{\dagger}_{j,\uparrow}\hat c_{j,\downarrow}+\hat c^{\dagger}_{j,\downarrow}\hat c_{j,\uparrow},\\
	\hat \sigma_{y}^{(g,j)}&=&-i(\hat c^{\dagger}_{j,\uparrow}\hat c_{j,\downarrow}-\hat c^{\dagger}_{j,\downarrow}\hat c_{j,\uparrow}),\\
	\hat \sigma_{z}^{(g,j)}&=&\hat c^{\dagger}_{j,\uparrow}\hat c_{j,\uparrow}-\hat c^{\dagger}_{j,\downarrow}\hat c_{j,\downarrow},\\
	\hat n^{(g,j)}&=&\hat c^{\dagger}_{j,\uparrow}\hat c_{j,\uparrow}+\hat c^{\dagger}_{j,\downarrow}\hat c_{j,\downarrow}.
\end{eqnarray}
In  addition,  $\hat\sigma_x^{(e,\alpha)}=|\uparrow\rangle_{\alpha}\langle\downarrow|+|\downarrow\rangle_{\alpha}\langle\uparrow|$ ,
$\hat\sigma_y^{(e,\alpha)}=-i(|\uparrow\rangle_{\alpha}\langle\downarrow|-|\downarrow\rangle_{\alpha}\langle\uparrow|)$, and $\hat \sigma_{z}^{(e,\alpha)}=|\uparrow\rangle_{\alpha}\langle\uparrow|-|\downarrow\rangle_{\alpha}\langle\downarrow|$ ($\alpha=1,...,N_e$)  are the components of the Pauli-operator vectors  $\hat{\bm \sigma}^{(e,\alpha)}$  of  the  $\alpha$-th  $e$-atom, which is fixed at $s_\alpha$-th site.

In summary, as shown in  Fig.~\ref{scheme2}(d), using our approach one can realize a system described by the time-independent 1D lattice model 
(\ref{hmb1}), where the $g$-atoms hop between nearest sites while the $e$-atoms are located at particular sites.  
Furthermore, one can tune the effective hopping rate $J_{\rm eff}$ of the $g$-atom and the effective interaction between the $g$- and $e$-atoms by changing the  control parameters $\{\delta J,\nu\}$, as shown in Eq.~(\ref{jueff}, \ref{Omega}).


\subsection{Example of Experimental Realization}
\label{exp}

In the end of this section, we show an example of the experimental realization of our scheme via 
ultracold $^{171}\mathrm{Yb}$ atoms. 
As shown in Sec.~\ref{setup}, the basic setup of our scheme includes a 
tilted quasi-1D confinement along the $z$-direction (i.e.,
nuclear-spin independent harmonic confinement in the $x$-$y$ plane), and 
a nuclear-spin dependent lattice along the $z$-direction.
For $^{171}\mathrm{Yb}$ atoms, this confinement and lattice
can
be realized via laser beams with magic wavelength $\lambda_\perp=759.4\text{nm}$ and 
another wavelength $\lambda_z$, respectively.
Consequently, the entire potential energy experienced by  the $j$-atom ($j=g,e$) is given by
\begin{eqnarray}
	V^{(j)}&=&V_\perp\sin^2\big(k_\perp\sqrt{x^2+y^2}\big)+V_z^{(j)}\sin^2(k_z z)\nonumber\\
	&&+mg(z\sin\theta+x\cos\theta),\nonumber\\
	&&\hspace{4.5cm}(j=g,e),
\end{eqnarray}
where $k_{\perp}=2\pi/\lambda_\perp$, $k_z=2\pi/\lambda_z$, $m$ is the single-atom mass, $g$ is the gravitational acceleration.  As shown in Fig.~\ref{scheme2}(a), the $y$-axis is 
always perpendicular to the gravity direction, and the angle between $z$-direction and 
the horizontal direction 
 is $\theta$.
Moreover, the ratio $V_z^{(e)}/V_z^{(g)}$ 
between the depth of the 1D lattice for $e$- and $g$-atoms
is determined by $\lambda_z$. Here we choose $\lambda_z=670\text{nm}$, so that 
${V_z^{(e)}}/{V_z^{(g)}}\approx 3.3$.
Additionally, in our scheme $V_\perp$ is time-independent, while $V_z^{(g)}$ oscillates with time with a small amplitude. Specifically, we have 
\begin{eqnarray}
	V_z^{(g)}(t)= V_{z0}^{(g)} +\delta V_z^{(g)}\sum_{l=1}^3 \bigg(e^{i\omega_lt}+ e^{-i\omega_lt}\bigg),\label{V_zgt}
\end{eqnarray}
where $\omega_{1,2,3}$ are functions of  parameters $\Delta$ and $\nu$, as shown in  Eq.~(\ref{fre}).
Due to the aforementioned relation ${V_z^{(e)}}/{V_z^{(g)}}\approx 3.3$, the intensity $V_z^{(e)}$ also slightly oscillates with time.

\begin{table}[t]
	\includegraphics[width=0.95\linewidth]{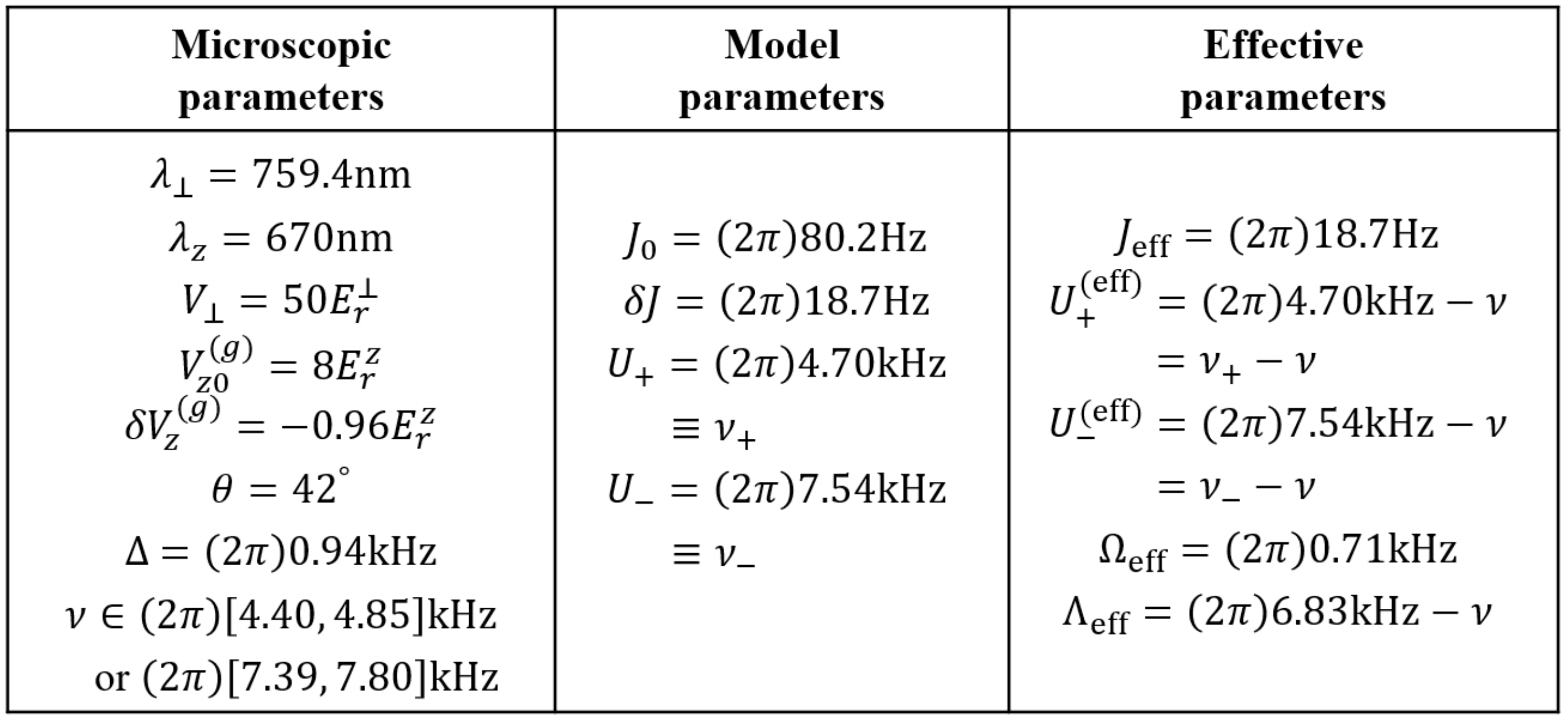}
	\caption{
	Representative parameters for implementing our scheme with $^{171}$Yb atoms. Here $E_r^z=\hbar^2 k_z^2/(2m)$ is the recoil energy of the $g$-atom along $z$, while $E_r^\perp=\hbar^2 k_\perp^2/(2m)$ is the recoil energy common to both $g$- and $e$-atoms in the $x$-$y$ plane. The wave numbers are defined as $k_z=2\pi/\lambda_z$ and $k_\perp=2\pi/\lambda_\perp$.
	}
	\label{table1}
\end{table}

When the microscopic parameters $(V_\perp,$ $V_{z0}^{(g)}$, $ \delta V_z^{(g)}$, $\theta, \Delta,\nu)$ 
are tuned appropriate regions, the time-dependent 1D model $\hat H(t)$ of Eq.~(\ref{Ht}), with $J(t)$ being given by Eq.~(\ref{Jt}), can be realized via our setup.
The detailed derivation of the model parameters $(J_0,\delta J, \Delta, U_{\pm })$ of the Hamiltonian $\hat H(t)$  from the microscopic parameters is given  
in appendix~\ref{J_and_U}. In Table~\ref{table1} we provide the result for a typical case.

\begin{figure}[t]
	\centering
	
	\begin{overpic}[width=0.9\linewidth]{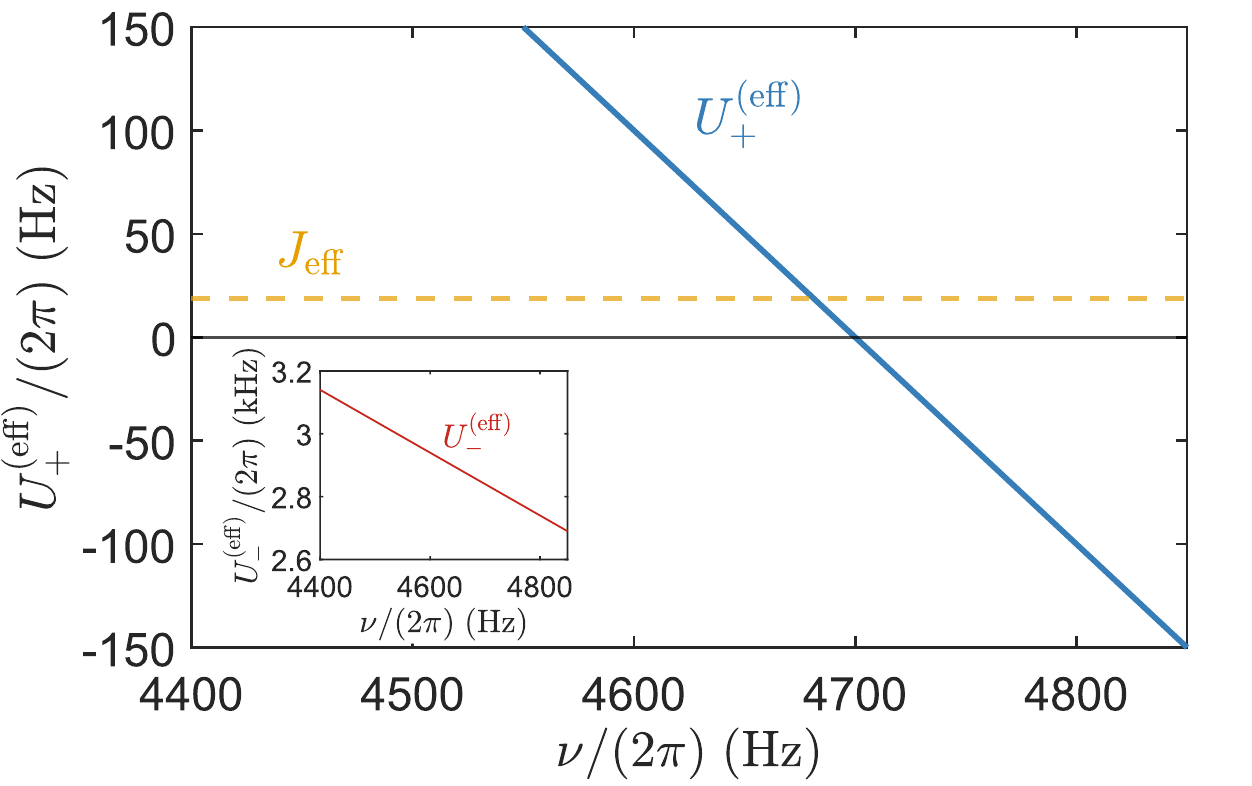}
		\put(-1,62){\bfseries (a)}
	\end{overpic}
	
	\vspace{1.5em}
	
	\begin{overpic}[width=0.9\linewidth]{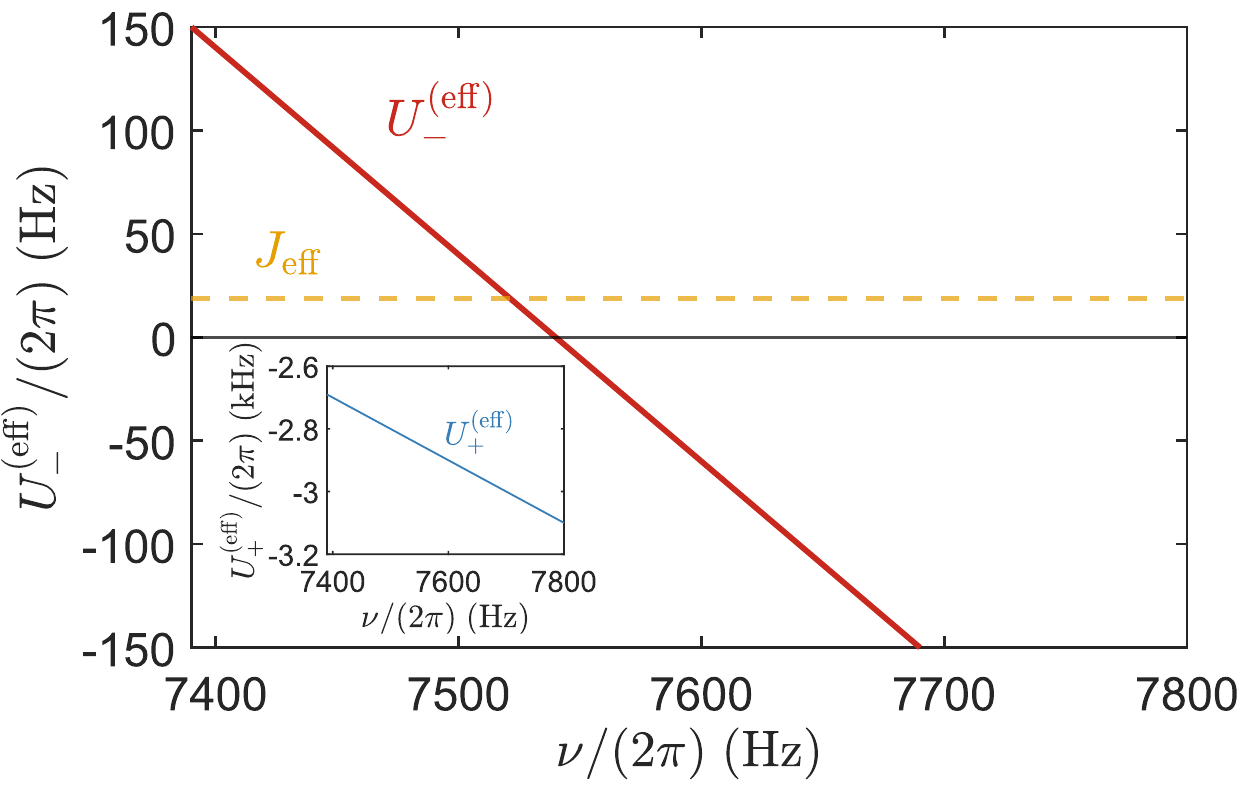}
		\put(-1,62){\bfseries (b)}
	\end{overpic}

	\caption{\textbf{(a):} $U_{+}^{\rm (eff)}$ (blue solid line) as a function of $\nu$
for $\nu/(2\pi)\in[4.40,4.85]\,\mathrm{kHz}$.
\textbf{(b):} $U_{-}^{\rm (eff)}$ (red solid line) as a function of $\nu$
for $\nu/(2\pi)\in[7.39,7.80]\,\mathrm{kHz}$.
The parameters are listed in Table~\ref{table1}.
For comparison, the yellow dashed line in each panel indicates the
corresponding value of $J_{\rm eff}$, which is independent of $\nu$.
The insets show $U_{-}^{\rm (eff)}$ and $U_{+}^{\rm (eff)}$
for the cases shown in (a) and (b), respectively.}
	\label{Ueff_fig}
\end{figure}

Furthermore, the condition (\ref{c3}) is satisfied for the case shown in Table~\ref{table1}. Consequently, our system can be approximately  described by the effective Hamiltonian  $\hat H_{\rm eff}$ in Eq.~(\ref{heff1}) or Eq.~(\ref{heff2}).
Table~\ref{table1} we show the
 parameters $(J_{\rm eff}, U^{\rm (eff)}_{\pm})$ or 
$(J_{\rm eff}, \Omega_{\rm eff}, \Lambda_{\rm eff})$ of the  effective Hamiltonian.
In Fig.~\ref{Ueff_fig} we further illustrate the dependence of $U^{\rm (eff)}_{\pm}$ on the parameter $\nu$. It has been shown that varying $\nu$ allows $U^{\rm (eff)}_{\pm}$
to be tuned to values smaller than, equal to, or larger
than $J_{\rm eff}$. Consequently, $\Lambda_{\rm eff}$ and
$\Omega_{\rm eff}$ can be tuned over a wide range.


\section{Spin-Exchanging Scattering}
\label{scattering}

Using our above approach for the manipulation of SEI, one can realize and control various  SEI related physical effects. In this section we illustrate an example.
We consider a two-atom system  described by the effective 1D lattice model with Hamiltonian $\hat H_{\rm eff}$ of Eq.~(\ref{heff1}). Specifically, the $g$-atom hops in the lattice, and  the $e$-atom is localized at the $0$-th site. The inter-atomic SEI is described by the 
intensities $U^{\rm (eff)}_{\pm}$. 

\begin{figure}[t]
	\centering
	\includegraphics[width=0.95\linewidth]{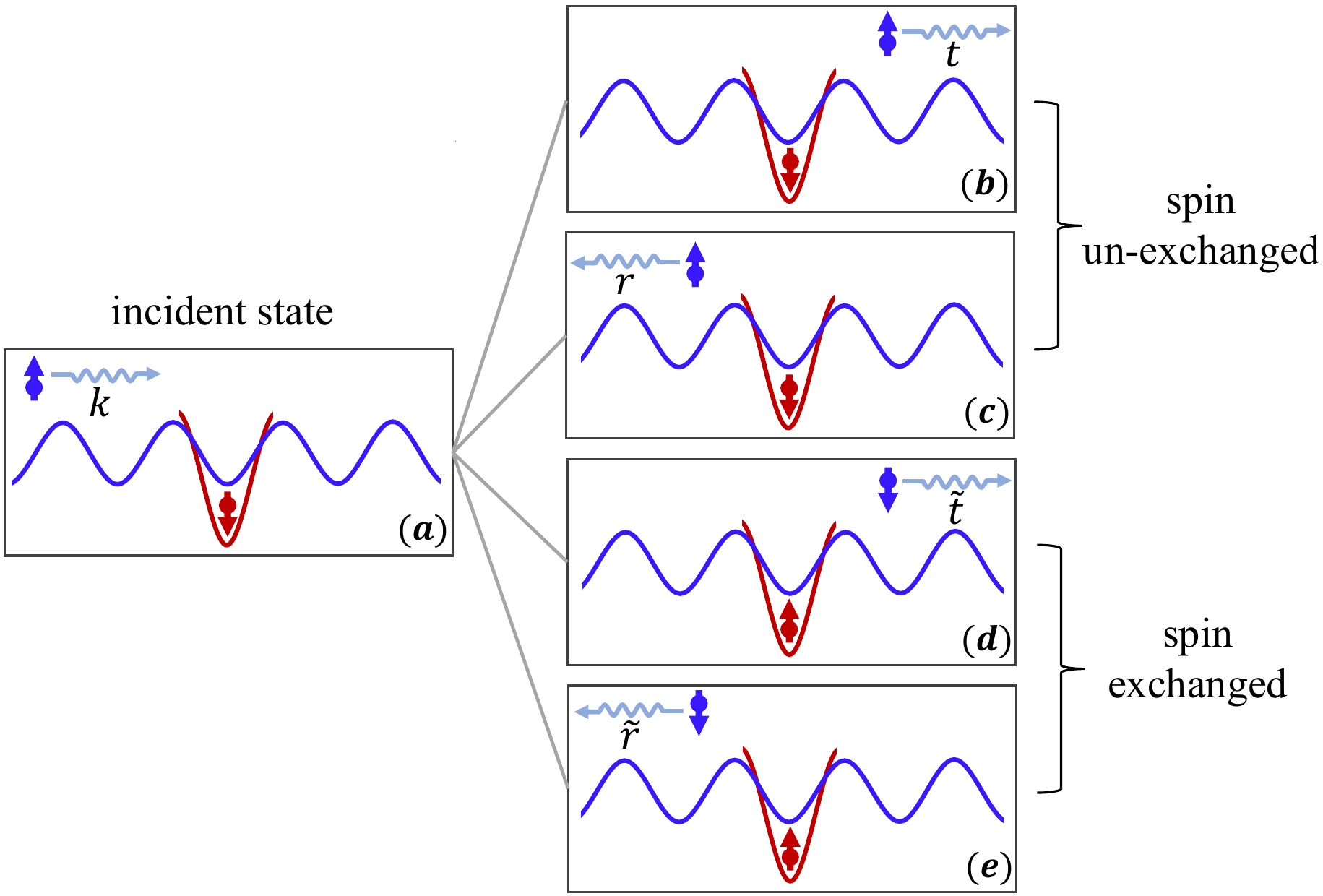}
	\caption{Schematic illustration of the scattering processes between an incident  $g$-atom and a localized $e$-atom. For details, see Sec.~\ref{scattering}.}
	\label{scatterfig}
\end{figure}

We consider the scattering process shown in Fig.~\ref{scatterfig}. Specifically, $g$- and $e$-atoms are initially in spin states 
$|\uparrow\rangle_g$ and $|\downarrow\rangle_e$, respectively, and the $g$-atom 
is incident from the left with momentum $k$ (Fig.~\ref{scatterfig}(a)). After the scattering process, the spin configuration of the two atoms can either remain unchanged (Fig.~\ref{scatterfig}(b, c)) or be exchanged, resulting in the state 
$|\downarrow\rangle_g|\uparrow\rangle_e$ (Fig.~\ref{scatterfig}(d, e)). Meanwhile, the $g$-atom can either transmit through the location of the $e$-atom (Fig.~\ref{scatterfig}(b, d)) or be reflected by the $e$-atom (Fig.~\ref{scatterfig}(c, e)).

Furthermore, this scattering process can be described by 
the scattering state $|\Psi_k\rangle$, which satisfies 
\begin{eqnarray}
\hat H_{\rm eff}|\Psi_k\rangle=E_k|\Psi_k\rangle,\label{se}
\end{eqnarray}
with $E_k=-2J_{\rm eff}\cos k$ ($k\in [0,\pi]$), and 
can be expressed as
\begin{eqnarray}
	|\Psi_k\rangle
	=
	\sum_j \psi_j^{(\uparrow\downarrow)} |j\rangle |\uparrow\rangle_g |\downarrow\rangle_e
	+
	\sum_j \psi_j^{(\downarrow\uparrow)} |j\rangle |\downarrow\rangle_g |\uparrow\rangle_e,\nonumber\\
	\label{psik}
\end{eqnarray}
with
\begin{eqnarray}
	\psi_j^{(\uparrow\downarrow)}
	&=&
	\begin{cases}
		e^{ikj}+r\,e^{-ikj}, & j\le0\\[5pt]
		t\,e^{ikj}, & j>0
	\end{cases},
	\label{C_updown}
	\\[10pt]
	\psi_j^{(\downarrow\uparrow)}
	&=&
	\begin{cases}
		\tilde r\,e^{-ikj}, & j\le0\\[5pt]
		\tilde t\,e^{ikj}, & j>0
	\end{cases}.
	\label{C_downup}
\end{eqnarray}
Here $r$ and $t$ are the reflection and transmission amplitudes without spin exchange, respectively (Fig.~\ref{scatterfig}(b, c)), while $\tilde r$ and $\tilde t$ are the reflection and transmission amplitudes  with spin exchange, respectively (Fig.~\ref{scatterfig}(d, e)).  
By solving the Schr\"odinger equation (\ref{se}) under conditions of Eq.~(\ref{psik}-\ref{C_downup}), we analytically derive all the four amplitudes $r, t, \tilde r, \tilde t$ (see Appendix~\ref{b1} for details).  
Here we focus on the total spin-exchange probability $P_{\mathrm{exc}}(k)$ of this scattering process, i.e., the total probability of reflection and transmission   with spin exchange:
\begin{eqnarray}
	P_{\mathrm{exc}}(k)
	=
	|\tilde r|^2 + |\tilde t|^2,
	\label{PSE}
\end{eqnarray}
and focus on the example with parameters given in Table~\ref{table1}.

We first consider the case that  the SEI is not tuned via our approach (i.e., $\delta J=0$). In this case the system is governed by the Hamiltonian in Eq.~(\ref{Ht}), with $J(t)=J_0$ and $\Delta=0$, and the values of $J_0$ and $U_\pm$ being given in Table~\ref{table1}. With straightforward calculations we find that 
the spin-exchange probability $P_{\mathrm{exc}}(k)$ is of the order of $10^{-4}$ or smaller. Therefore, the spin-exchange effect in this system is negligible.

\begin{figure}[t]
	\centering
	
	\begin{overpic}[width=0.75\linewidth]{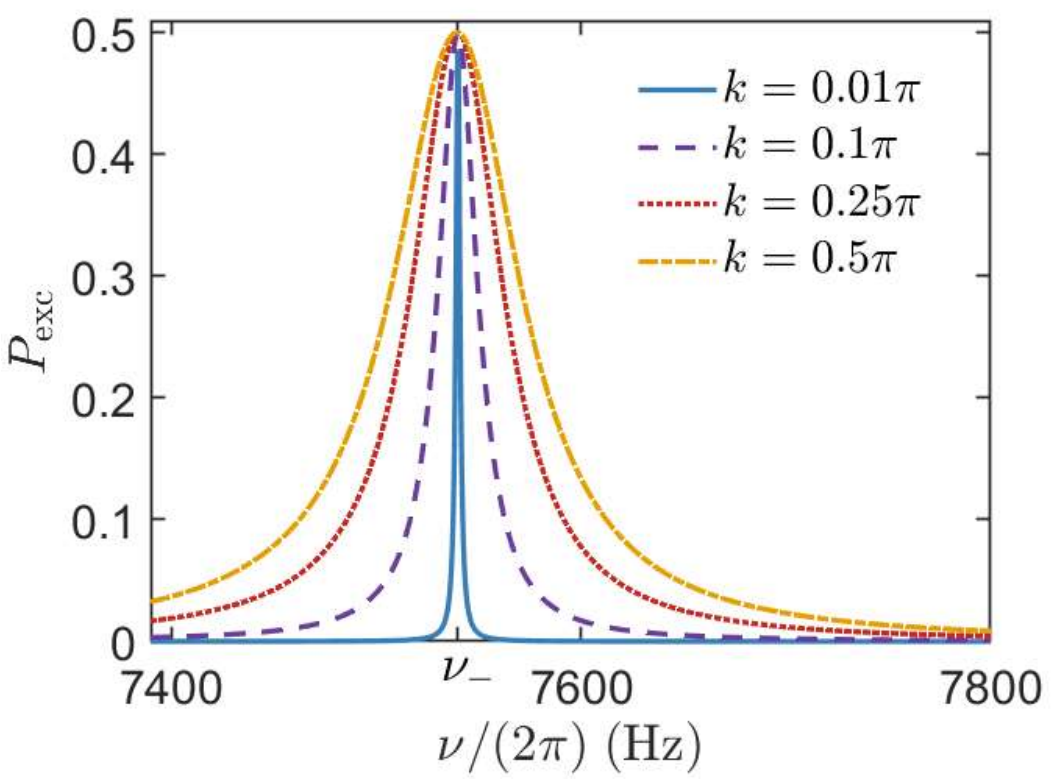}
		\put(-3,72){\bfseries (a)}
	\end{overpic}
	
	\vspace{1.5em}
	
	\begin{overpic}[width=0.75\linewidth]{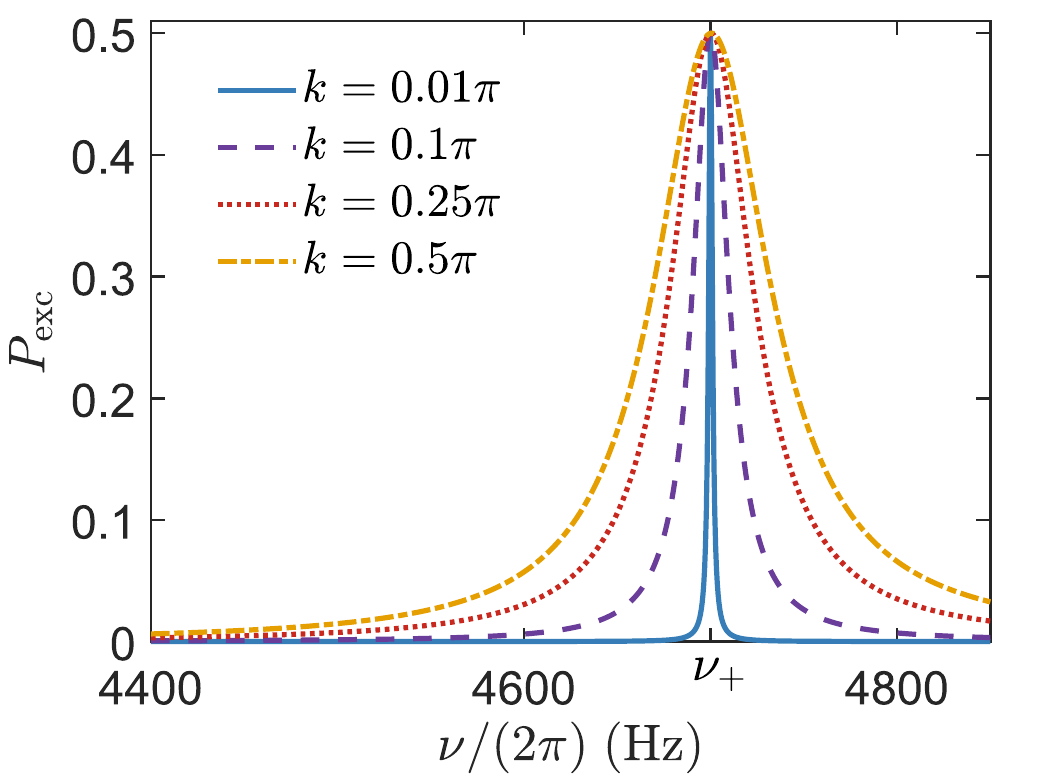}
		\put(-3,72){\bfseries (b)}
	\end{overpic}

	\caption{Total spin-exchange probability $P_{\rm exc}$ as a function of the modulation frequency $\nu$ for different incoming wave numbers $k=0.01\pi$ (blue solid line), $0.1\pi$ (purple dashed line), $0.25\pi$ (red dotted line), and $0.5\pi$ (yellow dash-dotted line). Here we show the results for cases with parameters in Table~\ref{table1}, with $\nu$ being around $\nu_+\equiv (2\pi)4.70{\rm kHz}$ ((a)), and $\nu_-\equiv(2\pi)7.54{\rm kHz}$ (b).}
	\label{sex}
\end{figure}

When our control approach is applied, the system is governed by  the effective Hamiltonian $\hat H_{\rm eff}$ in (\ref{heff1}), with $J_{\rm eff}=(2\pi)18.7$Hz,  and
$U_\pm^{(\rm eff)} = \nu_\pm-\nu$. Here the frequency $\nu$ is the control parameter, as shown in the above sections,
while $\nu_+\equiv (2\pi)4.70{\rm kHz}$, and $\nu_-\equiv(2\pi)7.54{\rm kHz}$ (Table~\ref{table1}). Therefore, by changing the frequency $\nu$, one can tune the value of $U_\pm^{(\rm eff)}$, and thus control the  spin-exchange probability $P_{\mathrm{exc}}(k)$.

To illustrate this control effect, 
in Fig.~\ref{sex}(a) and (b) we show $P_{\mathrm{exc}}(k)$ as a function of $\nu$, with respective various incident momentum $k$, for cases with $\nu$ being around $\nu_+$ and $\nu_-$, resepctively. It is shown that the spin-exchange  
probability $P_{\mathrm{exc}}(k)$ takes the maximum value $P_{\mathrm{exc}}^{\rm (max)}(k)$, which is almost 0.5,
 when the control frequency $\nu$ is tuned to $\nu=\nu_+$ or $\nu=\nu_-$, i.e., when 
 $U_+^{(\rm eff)}$ or $U_-^{(\rm eff)}$ are tuned to zero.
Moreover, 
$P_{\mathrm{exc}}(k)$ is resonantly enhanced
in the region around $\nu=\nu_\pm$, and is close to zero outside these resonance regions.
As shown in the detailed analysis of Appendix~\ref{b2}, these results can be explained via the quantum interference between the two independent scattering channels of our system, which corresponds to the spin singlet and triplet states of the atoms.

The above results clearly show that using our modulation approach  one can significantly enhance and effectively control the spin-exchange probability $P_{\mathrm{exc}}(k)$, by changing the control frequency $\nu$. This is a typical example for the possible observable effects realized by our approach.


\section{Summary}
\label{summary}

In this work, we propose a scheme for controlling the SEI between ultracold alkaline-earth atoms in the $g$ and $e$ states through periodic modulation of the optical-lattice depth. Unlike approaches based on CIRs, our scheme does not require the interatomic scattering length to be comparable to the characteristic confinement length. The experimental parameters required to implement the scheme are estimated for $^{171}$Yb atoms. We further illustrate an example of physical consequences of this SEI control by analyzing two-body spin-exchange scattering. Our scheme provides a useful tool for quantum simulations of SEI-induced many-body phenomena, such as the Kondo effect.


\begin{acknowledgments}
 This work was supported by  the National Key Research and Development Program of China (Grant No.2022YFA1405300) (P. Zhang), the Innovation Program for Quantum Science and Technology (Grant No. 2023ZD0300700) (P. Zhang), and the Outstanding Innovative Talents 
 Cultivation Funded Programs 2026 of Renmin University of 
 China (C. Xu). 
\end{acknowledgments}

\appendix
\begin{widetext}

\section{Conditions in Eq.~(\ref{c3})}
\label{condition}

In this appendix, we explain why the conditions listed in Eq.~(\ref{c3}) allow us to neglect the oscillating terms in the rotating-frame Hamiltonian $\hat H_R(t)$ of Eq.~(\ref{hrt}), leaving the time-independent effective Hamiltonian $\hat H_{\rm eff}$ given in Eq.~(\ref{heff1}).
In the following, we first show a general sufficient condition under which the oscillating terms in a Hamiltonian can be ignored (Sec.~\ref{prin}), and then apply this 
sufficient condition to our system (Sec.~\ref{ous}).

\subsection{General Sufficient Condition for Neglecting Rapidly Oscillating Terms}
\label{prin}

Consider a quantum system with a  time-dependent Hamiltonian 
\begin{eqnarray}
	\hat h(t)
	=
	\sum_{i=1}^n \left(
	\hat h_i e^{-i\omega_i t}
	+ h.c.
	\right)
	+ \hat h_0(t),
	\label{a1}
\end{eqnarray}
where the frequencies $\omega_i$ are positive and the operators $\hat h_i$ are time independent for $i=1,\ldots,n$. The term $\hat h_0(t)$ may itself depend on time. We define the amplitude $\|\hat h_\alpha\|$ of $\hat h_\alpha$ ($\alpha=0,\ldots,n$)  as the maximum absolute difference between any two of its eigenvalues.

For this system, one can neglect the terms $\sum_{i=1}^n \left(
	\hat h_i e^{-i\omega_i t}
	+ h.c.
	\right)$, and make the approximation
\begin{eqnarray}
\hat h \approx \hat h_0(t),\label{negro}
\end{eqnarray}
under the condition
\begin{eqnarray}
\|\hat h_{\alpha}\|\ll\omega_{1,...,n}, \ \ {\rm for\  all}\ \  \alpha=0,1,...,n.
\label{cond}
\end{eqnarray}
These inequalities require the amplitude of each term, including $\hat h_0(t)$, to be much smaller than every oscillation frequency. If $\hat h_0$ is time independent, Eq.~(\ref{negro}) gives a time-independent effective Hamiltonian. If $\hat h_0=0$, the effective Hamiltonian vanishes to this order. Notice that the inequalities in Eq.~(\ref{cond}) are {\it sufficient but not necessary} for the approximation in Eq.~(\ref{negro}).

\subsection{Application to Our System}
\label{ous}

Now we consider our system in the main text.
Substituting Eqs.~(\ref{Ht}) and (\ref{H0}) into Eq.~(\ref{hrt}) gives the Hamiltonian in the rotating frame (denoted as frame $R$),
\begin{eqnarray}
	\hat H_R(t)&=&-J(t)\sum_{j\ne 0,-1}\bigg(|j+1\rangle_g \langle j|e^{i\Delta t}+h.c.\bigg)\otimes \hat {\cal I}_{\rm spin}-J(t)\bigg(|j=0\rangle_g \langle j=-1|e^{i(\Delta+\nu) t}+h.c.\bigg)\otimes \hat {\cal I}_{\rm spin}
	\nonumber\\
&&	-J(t)\bigg(|j=1\rangle_g \langle j=0|e^{i(\Delta-\nu) t}+h.c.\bigg)\otimes \hat {\cal I}_{\rm spin}+H_{IR},
	\label{HR_all}
\end{eqnarray}
where time-independent interaction term is
\begin{eqnarray}
	\hat H_{IR}
	=
	|j=0\rangle_g\langle j=0|
	\otimes
	\left[
	(U_+-\nu)\hat{\cal P}_+
	+(U_--\nu)\hat{\cal P}_-
	\right].
\end{eqnarray}
The Hamiltonian in Eq.~(\ref{HR_all}) has the form of Eq.~(\ref{a1}). In the parameter regime of interest, however, the magnitudes of the interaction energies $U_\pm-\nu$ can be comparable to or larger than some of the oscillation frequencies. We therefore introduce a second rotating frame $R'$ in which the interaction energies enter the phases of the hopping terms. The state in this frame is
\begin{eqnarray}
	|\psi(t)\rangle_{R'}=e^{i\hat H_{IR}t}|\psi(t)\rangle_R.
	\label{newR}
\end{eqnarray}
The corresponding Hamiltonian is
\begin{eqnarray}
	\hat H_{R'}(t)&=&e^{i\hat H_{IR}t}\big[\hat H_R(t)-\hat H_{IR}\big]e^{-i\hat H_{IR}t}
	\nonumber\\
	&=&-J(t)\sum_{j\ne -1,0}\bigg(|j+1\rangle_g \langle j|e^{i\Delta t}+h.c.\bigg)\otimes \hat {\cal I}_{\rm spin}-J(t)\sum_{s=\pm}\bigg(|j=0\rangle_g \langle j=-1|e^{i(\Delta+U_s) t}+h.c.\bigg)\otimes \hat{\cal P}_s\nonumber\\
	&&-J(t)\sum_{s=\pm}\bigg(|j=1\rangle_g \langle j=0|e^{i(\Delta-U_s) t}+h.c.\bigg)
	\otimes \hat{\cal P}_s.
\end{eqnarray}
According to the expression Eq.~(\ref{Jt}) of $J(t)$, the amplitudes of every term of  $\hat H_{R'}(t)$ is $J_0$ or $\delta J$.
Thus, using the criterion in Sec.~\ref{prin}, we find that 
under the conditions in Eq.~(\ref{c3}), we can neglect several oscillating terms of $\hat H_{R'}(t)$, i.e., make the approximation
\begin{eqnarray}
	\hat H_{R'}(t)
	\approx\hat H_{R'}^{({\rm app})}(t)&:=&-\delta J\sum_{j\neq -1, 0}\bigg(|j+1\rangle_g \langle j|+h.c.\bigg)\otimes \hat {\cal I}_{\rm spin}-\delta J\sum_{s=\pm}\bigg(|j=0\rangle_g \langle j=-1|e^{i(U_s-\nu) t}+h.c.\bigg)\otimes \hat{\cal P}_s\nonumber\\
	&&-\delta J\sum_{s=\pm}\bigg(|j=1\rangle_g \langle j=0|e^{i(\nu-U_s) t}+h.c.\bigg)
	\otimes \hat{\cal P}_s.\label{hrpapp}
\end{eqnarray}
Transforming Eq.~(\ref{hrpapp}) back to frame $R$ gives the effective Hamiltonian $e^{-i\hat H_{IR}t}[\hat H_{R'}^{({\rm app})}(t)+\hat H_{IR}]e^{i\hat H_{IR}t}$. Direct calculation further yields that this effective Hamiltonian is just $\hat H^{\rm (eff)}$ in Eq.~(\ref{heff1}).

\section{Derivation of Model Parameters}
\label{J_and_U}

In this appendix we show our derivation from the microscopic parameters
$(V_\perp, V_{z0}^{(g)},  \delta V_z^{(g)}, \theta)$
 to the  model parameters $(J_0,\delta J, \Delta,  U_{\pm })$, for the example shown in 
 Sec.~\ref{exp}.
We first notice that, for both the $g$- and $e$-atom, the single-atom Hamiltonian in the $x$-$y$ plane is always given by
\begin{eqnarray}
\hat H_\perp&=&-\frac{\hbar^2}{2m}\frac{d^2}{dx^2}-\frac{\hbar^2}{2m}\frac{d^2}{dy^2}+
V_{xy}(x,y),
\end{eqnarray}
where the potential $V_{xy}(x,y)$ is 
\begin{eqnarray}
V_{xy}(x,y)=V_\perp\sin^2\big(k_\perp\sqrt{x^2+y^2}\big)+mgx\cos\theta.
\end{eqnarray}
Since $V_\perp$ is large, we expand $V_{xy}(x,y)$ around its minimum point, which is denoted as $(x_0,y_0)$, and keep the terms up to the second order of $(x-x_0)$ and $(y-y_0)$, i.e., approximate $V_{xy}(x,y)$ as a two-dimensional (2D) harmonic potential. We further assume that the transverse motion of both the $g$- and $e$-atoms are frozen at the ground state of $\hat H_\perp$, with wave function being denoted as $\phi_\perp(x,y)$. Under this approximation, our system can be described by a 1D model.

Furthermore, we calculate the 1D model parameters $(J_0,\delta J, \Delta)$ for the $g$-atom.
To this end, we express the Hamiltonian of the $g$-atom in the $z$-direction as
\begin{eqnarray}
\hat H_{z}^{(g)}=\hat H_{z0}^{(g)}+\hat H_{z1}^{(g)},
\end{eqnarray}
where
\begin{eqnarray}
\hat H_{z0}^{(g)}&=&-\frac{\hbar^2}{2m}\frac{d^2}{dz^2}+V_{z0}^{(g)}\sin^2(k_z z);\label{hz0}\\
\hat H_{z1}^{(g)}&=&\delta V_z^{(g)}\sum_{l=1}^3 \bigg(e^{i\omega_lt}+ e^{-i\omega_lt}\bigg)\sin^2(k_z z)+mgz\sin\theta.\label{hz1}
\end{eqnarray}
We make the single-band approximation, i.e., assume the atom is frozen in the lowest band 
of $\hat H_{z0}^{(g)}$. Additionally, we denote $|j\rangle_g$ ($j=0,\pm 1, \pm 2,....$)
as the $j$-th maximally localized Wannier state~\cite{Kohn1959PR}  of this band, with corresponding wave function $w_0(z-jl_z)$, where $l_z=\pi/k_z$. By expanding $\hat H_{z}^{(g)}$ with these Wannier states, we directly obtain the parameters $(J_0,\delta J, \Delta)$ of the model (\ref{Ht}). Specifically, we have
\begin{eqnarray}
	J_0	&=&
	\int dz\,
	w_0(z)
	\left[
	\frac{\hbar^2}{2m}\frac{d^2}{dz^2}
	-
	V_{z0}^{(g)}\sin^2(k_z z)
	\right]
	w_0(z+l_z)
	-\int dz\,
	w_0(z)
	\bigg[mgz\sin\theta
	\bigg]
	w_0(z+l_z);\\[8pt]
	\delta J
	&=&
	-
	\delta V_z^{(g)}
	\int dz\,
	w_0(z)
	\sin^2(k_z z)
	w_0(z+l_z);\\[8pt]
	\Delta
	&=&	mgl_z\sin\theta .
\end{eqnarray}
Notice that the second term in the expression for $J_0$ is negligible because its contribution to $J_0$ is smaller than $10^{-8}$.

Now we calculate the interaction strength $U_\pm$ between the $g$- and $e$-atoms in the model (\ref{Ht}). Notice that 
the Hamiltonian of the $e$-atom in the $z$-direction can be expressed as
$
\hat H_{z}^{(e)}=\hat H_{z0}^{(e)}+\hat H_{z1}^{(e)},
$
where $\hat H_{z0}^{(e)}$ and $\hat H_{z1}^{(e)}$
are still given by Eq.~(\ref{hz0}) and Eq.~(\ref{hz1}), respectively, with 
$V_{z0}^{(g)}$ and $\delta V_z^{(g)}$ in these equations being replaced by 
$V_{z0}^{(e)}$ and $\delta V_z^{(e)}$, respectively. As mentioned in our main text, we assume the $e$-atom is confined in the $0$-th site of the lattice potential in  $\hat H_{z0}^{(e)}$. Specifically, the wave function of the $e$-atom in the $z$-direction is
\begin{eqnarray}
	\phi_0^{(e)}(z)=
	\bigg(
	\frac{m\omega_z^{(e)}}{\pi\hbar}
	\bigg)^{1/4}
	\exp\bigg(
	-\frac{m\omega_z^{(e)} z^2}{2\hbar}
	\bigg).
\end{eqnarray}
where
$\omega_z^{(e)}=\big({2V_{z0}^{(e)}k_z^2}/{m}\big)^{\frac 12}$. Here we have approximated the lattice potential $V_{z0}^{(e)}\sin^2(k_z z)$ as a harmonic potential around the center point $z=0$ of the $0$-th site. Furthermore, we ignore the transition between $\phi_0^{(e)}(z)$ and the excited states of this harmonic potential, which is induced by $\hat H_{z1}^{(e)}$.
Additionally, the interaction between the $g$- and $e$-atoms can be described by a Huang-Yang pseudopotential \cite{HuangYang1957PR}:
\begin{equation}
	\hat U_{\rm HY}
	\equiv
	\frac{4\pi \hbar^2 }{m}
	\bigg[a_{eg}^{(+)} \hat {\cal P}_++a_{eg}^{(-)}\hat {\cal P}_-\bigg]
	\delta(\mathbf{r})
	\frac{\partial}{\partial r}(r \cdot),\label{uhy}
\end{equation}
where $\hat {\cal P}_\pm$ are the projection operators to the singlet/triplet spin states, as defined in Eq.~(\ref{proj}), $\bf r$ is the relative coordinate of the two atoms, and $a_{eg}^{(\pm)}$ are the scattering lengths of the two atoms in singlet/triplet states. For $^{171}$Yb atoms we have $a_{eg}^{(+)}=232(3)a_0$ and $a_{eg}^{(-)}=372(1)a_0$ \cite{Takahashi2019PRA, PengZhang2022PRR}, where $a_0$ is the Bohr radius.

To derive the interaction parameters $U_\pm$ for our 1D lattice model  (\ref{Ht}), we just project $\hat U_{\rm HY}$ on the state that both of the two atoms are in the $0$-th site, i.e., the state with wave function $\phi_\perp(x_e,y_e)\phi_\perp(x_g,y_g)\phi_0^{(e)}(z_e)w_0(z_g)$, where $(x_\alpha,y_\alpha,z_\alpha)$ ($\alpha=e,g$) is the coordinate of the $\alpha$-atom.
This projection directly leads to the interaction term $|j=0\rangle_g \langle j=0|\otimes(U_{+}\hat{\cal P}_++U_-\hat{\cal P}_-)$ in the model  (\ref{Ht}), with $U_\pm$ being given by
\begin{eqnarray}
	U_{\pm}
	&=&
	\frac{4\pi \hbar^2}{m}
	a_{eg}^{(\pm)}\int dxdydz |\phi_\perp(x,y)|^4
 |\phi_0^{(e)}(z)|^2
	|w_0(z)|^2.
\end{eqnarray}


\section{Calculation of the Reflection and Transmission Amplitudes}
\label{randt}

\subsection{Calculation of $P_{\mathrm{exc}}(k)$.}
\label{b1}

In this appendix we calculate the spin-exchange probability $P_{\mathrm{exc}}(k)$ for the scattering process in Sec.~\ref{scattering}, via 
solving the Schr\"odinger equation (\ref{se}) under conditions of Eq.~(\ref{psik}-\ref{C_downup}).
We first notice that the effective Hamiltonian $\hat H_{\rm eff}$ of Eq.~(\ref{heff1}) is diagonal in the basis of singlet state $|\rm S\rangle$ and triplet states $|{\rm T},m\rangle$ ($m=0,\pm 1$), which are defined in Eq.~(\ref{singlet}) and Eqs.~(\ref{t1}-\ref{t3}), respectively. Specifically, $\hat H_{\rm eff}$ can be re-expressed as 
\begin{eqnarray}
\hat H_{\rm eff}=\hat{\cal P}_+\hat H_{\rm eff}^{(+)}\hat{\cal P}_++\hat{\cal P}_-\hat H_{\rm eff}^{(-)}\hat{\cal P}_-,\label{nheff}
\end{eqnarray}
where $\hat{\cal P}_\pm$ are the projection operators for the singlet and triplet states, respectively, as defined in Eq.~(\ref{proj}), and $\hat H_{\rm eff}^{(\pm)}$ are the effective Hamiltonians for the singlet and triplet states, respectively, which are given by
\begin{eqnarray}
	\hat H_{\rm eff}^{(\pm)}
	=
	-J_{\mathrm{eff}}
	\sum_j
	\left(
	|j+1\rangle \langle j|
	+
	h.c.
	\right)
	+
	U_\pm^{\rm (eff)} |0\rangle \langle 0|.
	\label{Halpha}
\end{eqnarray}

Eq.~(\ref{nheff}) yields that in our problem there are two independent scattering channels, i.e., the 
channels with respect to singlet and triplet states. As a result, for a given incident state $k$, there are totally the following four independent scattering states:
\begin{eqnarray}
|\Psi_k^{(\rm S)}\rangle&=&\bigg(\sum_j \psi_j^{(+)} |j\rangle \bigg)|{\rm S} \rangle;\label{psiks}\\
|\Psi_k^{({\rm T},m)}\rangle&=&\bigg(\sum_j \psi_j^{(-)} |j\rangle \bigg)|{\rm T},m \rangle,\  (m=0,\pm 1).\label{psikt}
\end{eqnarray}
Notice that the three scattering states $|\Psi_k^{({\rm T},m)}\rangle$ ($m=0,\pm 1$), which correspond to the triplet spin states $|{\rm T},m\rangle$ ($m=0,\pm 1$), share the same spatial part $\sum_j \psi_j^{(-)} |j\rangle$. Furthermore, one can determine the spatial wave functions $\psi_j^{(\pm)}$ via solving the Schr\"odinger equations 
\begin{eqnarray}
\hat H_{\rm eff}^{(\pm)}\bigg(\sum_j \psi_j^{(\pm)} |j\rangle \bigg)=E_k\bigg(\sum_j \psi_j^{(\pm)} |j\rangle \bigg),
\end{eqnarray}
with $E_k=-2J_{\rm eff}\cos k$, under the scattering boundary conditions
\begin{eqnarray}
	\psi_j^{(\pm)}
	=
	\begin{cases}
		e^{ikj}+r_\pm e^{-ikj}, & j\leq 0\\[8pt]
		t_\pm e^{ikj}, & j>0
	\end{cases}\ \ .
	\label{psij}
\end{eqnarray}
Here $t_\pm$ and $r_\pm$ are the transmission and reflection amplitudes of the singlet/triplet channels, respectively. Straightforward calculations yields that 
\begin{eqnarray}
	t_\pm
	=
	\frac{2iJ_{\rm eff}\sin k}
	{2iJ_{\rm eff}\sin k-U_\pm^{\rm (eff)}},\ \ 
	r_\pm
	=
	\frac{U_\pm^{\rm (eff)}}
	{2iJ_{\rm eff}\sin k-U_\pm^{\rm (eff)}}.
	\label{rpm}
\end{eqnarray}

Now let us consider the scattering process studied in Sec.~\ref{scattering}, where in the incident state the $g$- and $e$-atoms are in the polarized state $|\uparrow\rangle_g|\downarrow\rangle_e$. Using the fact
\begin{eqnarray}
	|\uparrow\rangle_g|\downarrow\rangle_e
	=
	\frac{|{\rm T},0\rangle+|{\rm S}\rangle}{\sqrt{2}},
	\end{eqnarray}
we know that the scattering state $|\Psi_k\rangle$ corresponding to this scattering process is just a superposition of those in the singlet and triplet scattering channels, i.e.,
\begin{eqnarray}
|\Psi_k\rangle=\frac{1}{\sqrt 2}\bigg(|\Psi_k^{(\rm S)}\rangle+|\Psi_k^{({\rm T},0)}\rangle\bigg).\label{psikapp}
\end{eqnarray}
Comparing Eqs.~(\ref{psikapp}, \ref{psiks}, \ref{psikt}, \ref{psij}) with 
Eq.~(\ref{psik}-\ref{C_downup}) of Sec.~\ref{scattering}, and using the fact
\begin{eqnarray}
	|\downarrow\rangle_g|\uparrow\rangle_e
	=
	\frac{|{\rm T},0\rangle-|{\rm S}\rangle}{\sqrt{2}},
\end{eqnarray}
we finally obtain the transmission and reflection amplitudes with or without spin-exchange, i.e., $\{r,t;\tilde r,\tilde t\}$ in Eqs.~(\ref{C_updown}, \ref{C_downup}):
\begin{eqnarray}
	r
	&=&
	\frac{r_-+r_+}{2},\ \ \ \ \ 
	t=
	\frac{t_-+t_+}{2},\ \ \ \ \ 
	\tilde r
	=
	\frac{r_--r_+}{2},\ \ \ \ \ 
	\tilde t
	=
	\frac{t_--t_+}{2},\label{rtexp}
\end{eqnarray}
with $r_\pm$ and $t_\pm$ being given by Eq.~(\ref{rpm}).
Substituting these results into the expression $P_{\mathrm{exc}}(k)=
	|\tilde r|^2 + |\tilde t|^2$ (i.e., Eq.~(\ref{PSE})) of the spin-exchange probability, we directly obtain this probability.
	
\subsection{Explanation of Results of Fig.~\ref{sex}.}
\label{b2}

Now we explain the spin-exchange rate in the scattering process of the example of Sec.~\ref{scattering}, i.e., 
the results shown in Fig.~\ref{sex}. We first consider the cases that the control frequency $\nu$ is not close to $\nu_\pm$. In this case, according to the parameters shown in Table I, the interaction intensities $U_\pm^{\rm (eff)}$ are both much larger than the effective hopping rate $J_{\rm eff}$.
Equation~(\ref{rpm}) gives $r_\pm\approx -1$ and $t_\pm\approx 0$, indicating that the $g$-atom is reflected by the $e$-atom with nearly unit probability in both the singlet and triplet channels. Substituting this result into Eq.~(\ref{rtexp}), we immediately find that the spin-exchanging reflection and transmission amplitudes, i.e., $\tilde r$ and $\tilde t$, are all approximately zero, and thus we have $P_{\mathrm{exc}}(k)=
|\tilde r|^2 + |\tilde t|^2\approx 0$. The result for the case that SEI is not tuned via our approach can also be explained as above.

Now let us consider the case with $\nu=\nu_+$, where we have $U_+^{\rm (eff)}=0$ but $U_-^{\rm (eff)}\gg J_{\rm eff}$. According to Eq.~(\ref{rpm}), in this case we have $t_+=1$, $r_+=0$, $r_-\approx -1$ and $t_-\approx 0$, i.e., in the spin-singlet channel, the $g$-atom can completely transmit through the $e$-atom, while  in the spin-triplet channel, the $g$-atom is almost reflected. Using Eq.~(\ref{rtexp}), we 
find that  the spin-exchanging reflection and transmission amplitudes, i.e., $\tilde r$ and $\tilde t$, are both
approximately $-1/2$.
As a result, the spin-exchange rate is $P_{\mathrm{exc}}(k)=
|\tilde r|^2 + |\tilde t|^2\approx 1/2$, i.e., there is significant spin-exchanging effect in the scattering, as shown in Fig.~\ref{sex}(a). The result for the case with $\nu=\nu_-$ can also be explained with analysis similar as above.

Furthermore, the above analysis show that in our system the spin-exchanging scattering is enhanced when the scattering  amplitudes of singlet and triplet spin channels are quite different from each other, and is suppressed when the amplitudes of these channels are very similar. This is due to the fact that,
in our system the spin-exchanging scattering amplitudes $\tilde r$ and $\tilde t$ are determined by destructive interference of those for scattering in the singlet and triplet spin channels, as shown in 
Eq.~(\ref{rtexp}).

\end{widetext}

\end{document}